\documentclass[aps,prl,twocolumn,epsf,epsfig,amsmath]{revtex4}
\usepackage{latexsym}
\usepackage{hyperref}
\usepackage{times}
\usepackage{graphicx}
\usepackage{amsmath}
\usepackage{multirow} 
\usepackage{amsthm}
\usepackage{dcolumn}
\usepackage{bm}
\usepackage{ textcomp }
\usepackage{color}
\usepackage{amssymb}
\usepackage{xcolor}
\usepackage{easyReview}
\usepackage{mathdots}
\usepackage{float}
\usepackage{lineno}
\usepackage{todonotes}
\usepackage{array}

\newcommand{\beq}{\begin{eqnarray}}
\newcommand{\eeq}{\end{eqnarray}}

\begin{document}
\title{$\mathbb{Z}_2$ Universality of the Mott Transition}
\author{Jinchao Zhao$^{1,2}$}
\email{jinchao@ust.hk}
\author{Peizhi Mai$^{1}$, Gaurav Tenkila$^{1}$}         
\author{Philip W. Phillips$^{1}$}
\email{dimer@illinois.edu}

\affiliation{$^1$Department of Physics and Institute of Condensed Matter Theory, University of Illinois at Urbana-Champaign, Urbana, IL 61801, USA}
\affiliation{$^2$Department of Physics, Hong Kong University of Science and Technology, Clear Water Bay, Hong Kong, China}

\begin{abstract}

We demonstrate that the Mott transition exhibits universal scaling as a consequence of the breaking of a $\mathbb{Z}_2$ symmetry in momentum space.  A direct consequence of this discrete symmetry breaking is the charge or Mott gap itself.  From extensive numerics, we proffer that it is the charge compressibility that acts as the underlying order parameter as it is zero in the insulator and non-zero in the metallic state.  Additionally,  the Widom line (temperature of the extremum of the compressibility) obeys a universal scaling of $T_m=0.39U$ deep into the insulating state directly from $Z_2$ universality.  Furthermore, the temperature at which the second derivative of the compressibility has a minimum is independent of lattice geometry, exhibiting a universal scaling of $|U-U_c|^\alpha$ where $\alpha\approx 1$. Finally, our computational approach reproduces the key features of the doping dependence of the compressibility demonstrated in recent cold-atom quantum simulators of the Hubbard model, thereby corroborating our conclusions on $\mathbb{Z}_2$ universality. 

\end{abstract}
\date{October 2025}


\maketitle

Mott's proposal that a half-filled electronic band insulates without breaking any continuous symmetry is one of the most important claims in condensed matter physics. While simple mean-field (MF) ordering scenarios abound for the insulating state in a half-filled band, Mott\cite{mott} saw that such mechanisms are ultimately deficient because the observed gap of 0.6 eV in VO$_2$ is beyond any
energy scale entailed by dimerization of the vanadium
ions.  Similarly in the cuprates, a charge gap is present in the optical conductivity well above any temperature associated with Ne\'el ordering\cite{cooper,uchida}. Consequently, mean-field scenarios leave an explanatory residue, and the physics Mott had in mind (Mottness) must reside in the inherent strong correlations. That is, Mottness=MI-MF.

Attempts to get at Mottness have been largely directed at the Hubbard model either through state-of-the-art numerics\cite{dmft,Huangprr2022,tremblay,MaierRMP2005} or teasing out an order parameter\cite{castellani}.  Driven by Mott's original proposal that strong repulsions lead to single occupancy, Castellani, et al.\cite{castellani} proposed that double occupancy should serve as the order parameter for the Mott transition.  Consequently, the Mott transition should lie in the Ising universality\cite{castellani,dobro} class, which does have some experimental\cite{expmott1,expmott2} and phenomenological\cite{phillipsf} support. Theoretically, if double occupancy is the order parameter, then it should exhibit some sort of discontinuity across the Mott transition.  As the double occupancy is a first derivative of the free energy, the transition should be first order.  However, numerically\cite{dmft,dobro}, DMFT is the only method that yields a discontinuity of the double occupancy across the Mott transition in $d=\infty$ at a non-zero value of the on-site repulsion.  In fact, DMFT yields a density of states that has a central peak with no support on either side, consistent with a mixed metallic and paramagnetic insulating state.  Hence, in this scenario a first-order line terminates in a second-order critical endpoint. Contrastly, for the square lattice Hubbard model, state-of-the-art methods\cite{MaierRMP2005} reveal no critical value of $U$ for the insulating state to ensue and double occupancy\cite{double} is a simple smooth decreasing function of $U$.   Consequently, even the order of the transition is up for grabs.

If double occupancy is not the order parameter, lying in limbo then is a central claim of Mott physics that the transition lies in the Ising universality class.  It is this problem and the nature of the discontinuity across the Mott transition that we address in this note.  In a completely overlooked pamphlet, Anderson and Haldane (AH)\cite{haldane} showed that Fermi liquids possess a hidden $\mathbb{Z}_2$ symmetry. While we have used this symmetry argument extensively\cite{ppz2,ppfixedp}, we recount it here as it is central to our argument. The $\mathbb{Z}_2$ symmetry is in momentum space and represents the improper rotations of linear combinations of the Fermionic operators into one another. Simply, it maps 
creation operators of one spin species onto the annihilation counterpart,
$c_{p\uparrow}\rightarrow \pm c_{p\uparrow}^\dagger$ 
leaving the other spin species unscathed $c_{p\downarrow}\rightarrow c_{p\downarrow}$.  As the full Hamiltonian of a Fermi liquid has $O(4)$ symmetry, the $\mathbb{Z}_2$ arises as the quotient $O(4)/SO(4)=\mathbb{Z}_2$, where the $SO(4)$ arises from the equivalent $SU(2)$'s for the spin and the charge degrees of freedom.  This symmetry only holds for the locus of momenta at the Fermi level. The key point is that any non-Fermi liquid must arise simply by breaking the $\mathbb{Z}_2$ symmetry. Resultantly, any interaction involving a product of the occupancies $n_\uparrow n_\downarrow$ breaks the $\mathbb{Z}_2$ symmetry.  As AH\cite{haldane} did not provide any indices on the occupancies, there is a subtlety here. Not only must the operator break $\mathbb{Z}_2$ symmetry but it must also be a relevant perturbation to a Fermi liquid.  Keeping time-reversal intact, the simplest term that meets both requirements, as it has scaling dimension $-2$, is of the form $n_{p\uparrow}n_{p\downarrow}$.  It is easy to see that breaking $\mathbb{Z}_2$ with this relevant perturbation must produce upper and lower Hubbard bands. Writing the electron operator as $c_{p\uparrow}=c_{p\uparrow}(1-n_{p\downarrow})+c_{p\uparrow}n_{p\downarrow}=\xi_{p\uparrow}+\eta_{p\downarrow}$, we note that under $\mathbb{Z}_2$,
\begin{equation}
\left(
\begin{array}{l}
\zeta_{p\uparrow} \\[12pt]
\eta_{p\uparrow}
\end{array}
\right)
\xrightarrow{\mathbb{Z}_2}
\left(
\begin{array}{l}
\eta_{p\uparrow} \\[12pt]
\zeta_{p\uparrow}
\end{array}
\right),
\end{equation}
these two operators are interchanged, indicating a degeneracy.  This fails if $\mathbb{Z}_2$ is broken, resulting in the electron operator splinters into a lower and upper branch.  This is Mott's mechanism, laying plain that $\mathbb{Z}_2$ or the Ising universality class undergirds the Mott transition without invoking double occupancy as the order parameter. 

We develop this idea further and show that rather than double occupancy (see Appendix), it is the compressibility that exhibits a discontinuity across the Mott transition. We construct the Widom line\cite{stanley,dobro,tremblay} and show that it exhibits universality regardless of the locality of the underlying model, that is, real or momentum space. Hence, Mott physics can be classified via the universality of a second-order critical endpoint. By interpolating between $n_{p\uparrow}n_{p\downarrow}$ and the full Hubbard interaction simply by including momentum mixing\cite{twisted}, we find a universal trend in the compressibility, independent of the underlying interaction.  Our work complements prior high-temperature ($T>0.5t)$ compressibilities on the Hubbard model\cite{trivedi,trivedi2}.

We begin with the Hubbard model, which we write in momentum space
\begin{equation}
H_{\rm Hubb}= \sum_{k\sigma} \xi(k) n_{k\sigma}+\frac{U}{N}\sum_{{\bf k},{\bf p},{\bf q}\in \mathrm{BZ}} c^\dagger_{{\bf k}\uparrow} c_{{\bf k}-\bf q\uparrow}c^\dagger_{{\bf k}+{\bf p}\downarrow}c_{{\bf k}+{\bf p}+{\bf q}\downarrow},
\end{equation}
where $\xi_k=\epsilon_k-\mu$. The local-in momentum space Hatsugai-Kohmoto (HK)\cite{hk} model truncates the interaction to just a single momentum
\begin{equation}
H^{\rm HK}_{\rm int}=U\sum_{{\bf k}\in \mathrm{BZ}} c^\dagger_{{\bf k}\uparrow} c_{{\bf k}\uparrow}c^\dagger_{{\bf k}\downarrow}c_{{\bf k}\downarrow} \label{bHK}.
\end{equation}
which breaks the $\mathbb{Z}_2$ symmetry explicitly.  Elsewhere\cite{twisted} we showed in a procedure we termed the n-MMHK model how to interpolate between these models simply by including the momentum mixing the HK model lacks. Here $n$ is the number of mixed momenta.  At each point of the iteration scheme\cite{twisted}, which groups momenta into a larger cell and then hybridizes them, the interaction is still diagonal in the reduced Brillouin zone for the new cell. Consequently, the $\mathbb{Z}_2$ symmetry breaking is apparent at each grouping.  It is for this reason\cite{twisted} that our conclusions on the universality are independent of the momentum mixing and governed by the HK fixed point\cite{ppfixedp}.  

Since the HK model is tractable, we compute its compressibility first.   We are explicit as previous studies\cite{hk} have not emphasized the Widom line and have been limited to $d=1$.  As we will see, $d=1$ is peculiar and the generic behavior emerges only for $d\ge 2$.  
The total number of particles is given by
\begin{equation}
    N=\sum_{k\in \text{BZ}}n_k,
\end{equation}
where 
\begin{equation}
    n_k=\frac{1}{Z_k}\left(2e^{-\beta\xi_k}+2e^{-\beta(2\xi_k+U)}\right)
\end{equation}
with $Z_k=1+2e^{-\beta\xi_k}+e^{-\beta(2\xi_k+U)}$ the partition function.
Taking the derivative of the particle number $N$ with respect to the chemical potential $\mu$, 
\begin{equation}
   \chi=\sum_{k\in \text{BZ}}\frac{2\beta}{Z_k^2}\left(e^{-\beta\xi_k}+2e^{-\beta(2\xi_k+U)}+e^{-\beta(3\xi_k+U)}\right),
\label{eq:comp}
\end{equation}
results in the compressibility. 
\begin{figure}[t]
    \centering
    \includegraphics[width=\linewidth]{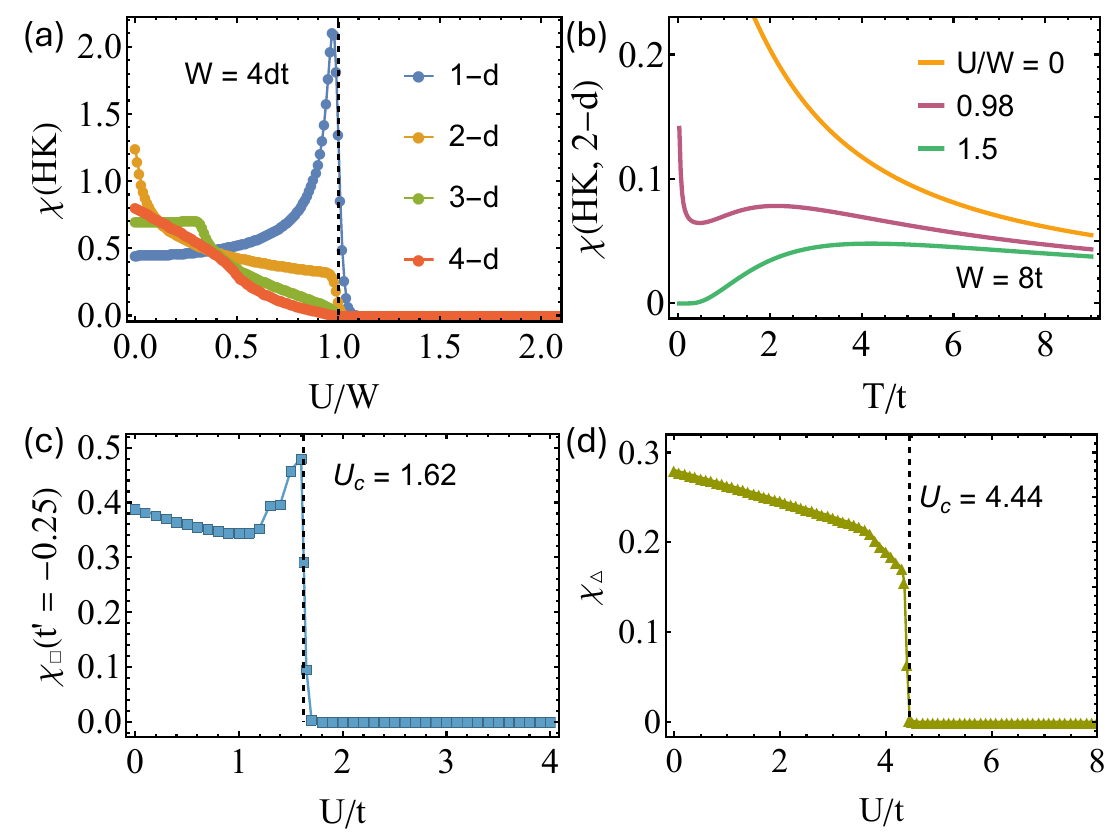}
    \caption{ (a) Low-temperature compressibility, $\chi$, as a function of $U/W$ in band HK model, where $W$ is the bandwidth, in $d=1$, $2$, $3$ and $4$. 
    (b) The compressibility as a function of temperature for a metal (yellow), an insulator (green), and a metal around the Mott transition (purple).
    (c) Low-temperature compressibility $\chi$ as a function of $U$in $4$-MMHK model on 2-d square lattice with $t'/t=-0.25$.
    (d) Low-temperature compressibility $\chi$ as a function of $U$in $4$-MMHK model on a 2-d triangular lattice. }
    \label{fig:chi_v_U}
\end{figure}
Also known as the thermal density of states, the compressibility corresponds to the total density of states at the filling surfaces. When the chemical potential lies in the gap, the zero density of states marks an insulator. From the low temperature compressibility as shown in Fig.~\ref{fig:chi_v_U}(a) for the band HK model in all dimensions, we find that the Mott metal-insulator transition always occurs at $U/W=1$ ($W$ is the non-interacting band width), where the upper Hubbard band is completely lifted from the lower Hubbard band. 

The $n$-momentum-mixing HK ($n$-MMHK) model introduces quantum fluctuation or non-trivial dynamics into the original band HK model by adding $n$ momentum scattering 
\begin{equation}
\begin{split}
    H^{\text{MMHK}}_{n}&=\frac{U}{n}\sum_{\mathbf{k}\in \rm{BZ}}\sum_{\mathbf{p}\in B_n}\sum_{\mathbf{q}\in B_n}c_{\mathbf{k}\uparrow}^\dagger c_{\mathbf{k-q}\uparrow}c_{\mathbf{k+p}\downarrow}^\dagger c_{\mathbf{k+p+q}\downarrow}\\
    &=U\sum_{\mathbf{k}\in \rm{rBZ}_n}\sum_\alpha n_{\mathbf{k}\alpha\uparrow}n_{\mathbf{k}\alpha\downarrow},
\end{split}
\end{equation}
systematically, where $\rm{rBZ}_n$ is the reduced Brillouin zone of the $n$-site (mixed momenta) unit cell, and $B_n$ is the set of reciprocal lattice points of the $n$-site unit cell that live in the first Brillouin zone of the original lattice as discussed in \cite{twisted}. In the last line, we changed to the orbital basis to switch to factorized blocks in $\rm{rBZ}_n$.

While introducing a few momentum scattering already captures much of Hubbard dynamics, the $n$-MMHK model remains exactly\cite{twisted} solvable. As shown in Fig.~\ref{fig:chi_v_U}(c), the Mott gap opens for $U>U_c\approx1.62$ in the $4$-MMHK model on a square lattice with nearest and next-nearest neighbor hopping $t$ and $t'$. Similarly, the $4$-MMHK model on the triangular lattice also has a finite $U_c\approx4.44$ as shown in Fig.~\ref{fig:chi_v_U}(d). 

The explicit dependence of the compressibility on the non-interacting density of states (See Appendix) is no longer valid in MMHK models. However, the low temperature compressibility still vanishes for $U>U_c$, marking the opening of a charge gap and the universality of the Mott transition in the family of HK-like models.  For all of these systems, the compressibility functions as the order parameter as it vanishes in the insulator and turns on in the metal.  As $\chi$ corresponds to a second derivative of the free energy, the underlying transition is second order with a universality of $\mathbb{Z}_2$.

We now consider the high-temperature regime where thermal excitation can play a factor at the metal-insulator transition. Physically, it is the vanishing density of states in the gap that leads to a concomitantly small value of $\chi$. On the other hand, thermal excitations can enhance the compressibility.  At sufficiently high temperature, the dominant energy scale is replaced by $k_BT$. In this regime, the insulator ``melts'' as its compressibility decreases if the temperature were to increase further, just as in a metal. In the infinite temperature limit, the particle number becomes irrelevant to the chemical potential, which means that $\chi\rightarrow 0$. Because $\chi$ is a positive-definite function, $\chi$ must have a maximum at some temperature, namely the melting temperature $T_m$. Fig.~\ref{fig:chi_v_U}(b) shows the typical compressibility as a function of temperature for a normal metal ($U=0$) and an insulator ($U>W$). We recognize the positive correlation between compressibility and temperature as the signature of insulating phases. At high enough temperature, a negative correlation between compressibility and temperature is expected for the melted insulator.

To introduce the universality of Mottness, we note that for a large enough value of $U$, the kinetic energy can be ignored. We simplify the analysis by only considering the particle-hole symmetric case and specialize to half-filling by setting $\mu=U/2$.  This is equivalent to switching to the atomic limit of the Hubbard model. The compressibility then becomes
\begin{equation}
\begin{split}
    \chi\rightarrow\chi^{\text{atomic limit}}&=\frac{\beta}{1+e^{\beta U/2}}.
\end{split}
\end{equation}
By taking the temperature derivative, we find
\begin{equation}
\begin{split}
    \frac{\partial\chi^{\text{atomic limit}}}{\partial \beta}&=\frac{2+e^{\beta U/2}(2-\beta U)}{2\left(1+e^{\beta U/2}\right)^2}.
\end{split}
\end{equation}
The melting temperature is thus located at $2+e^{\beta_m U/2}(2-\beta_m U)=0$, which yields $ \beta_m U=U/T_m\approx2.56$ or equivalently $T_m\approx0.39U$.  What we show here is that this value is universal and not determined at all by the kinetic energy. Rather, it arises from the general interplay between thermal fluctuations and strong correlations.  The kinetic energy is irrelevant as it remains small relative to either of these quantities.   For the case of Hubbard interactions, the large $U$ limit generates the identical atomic limit. This is the origin of the $\mathbb{Z}_2$ universality of the Mott transition.

To put the universality of the Mott insulator in context, we review what obtains in a band insulator. Consider the case where the gap size $\Delta\gg W$, with $W$ the bandwidth.  In this case, the kinetic energy can be ignored. We also set the chemical potential to the middle of the gap to keep the particle number fixed.  In this limit, the compressibility is simple:
\begin{equation}
    \chi^\text{band insulator}=\frac{4\beta}{2+e^{\beta\Delta/2}+e^{-\beta\Delta/2}}.
\end{equation}
The temperature derivative of the compressibility is
\begin{equation}
\begin{split}
   \frac{\partial\chi^\text{band insulator}}{\partial \beta}&=\frac{2e^{\beta \Delta/2}\left(2+\beta \Delta+e^{\beta \Delta/2}(2-\beta \Delta)\right)}{\left(1+e^{\beta \Delta/2}\right)^3}.
\end{split}
\end{equation}
The maximum compressibility is located at the solution to $2+\beta_m \Delta+e^{\beta_m \Delta/2}(2-\beta_m \Delta)=0$, which gives $\beta_m\Delta=\Delta/T_m\approx3.09$, or equivalently $T_m\approx0.32\Delta$. Note that the numerical prefactor is distinct from the Mott case, thereby serving as a further demarcation between Mott and band insulators.  Namely, they melt at fundamentally different temperatures.

\begin{figure}[t]
    \centering
    \includegraphics[width=\linewidth]{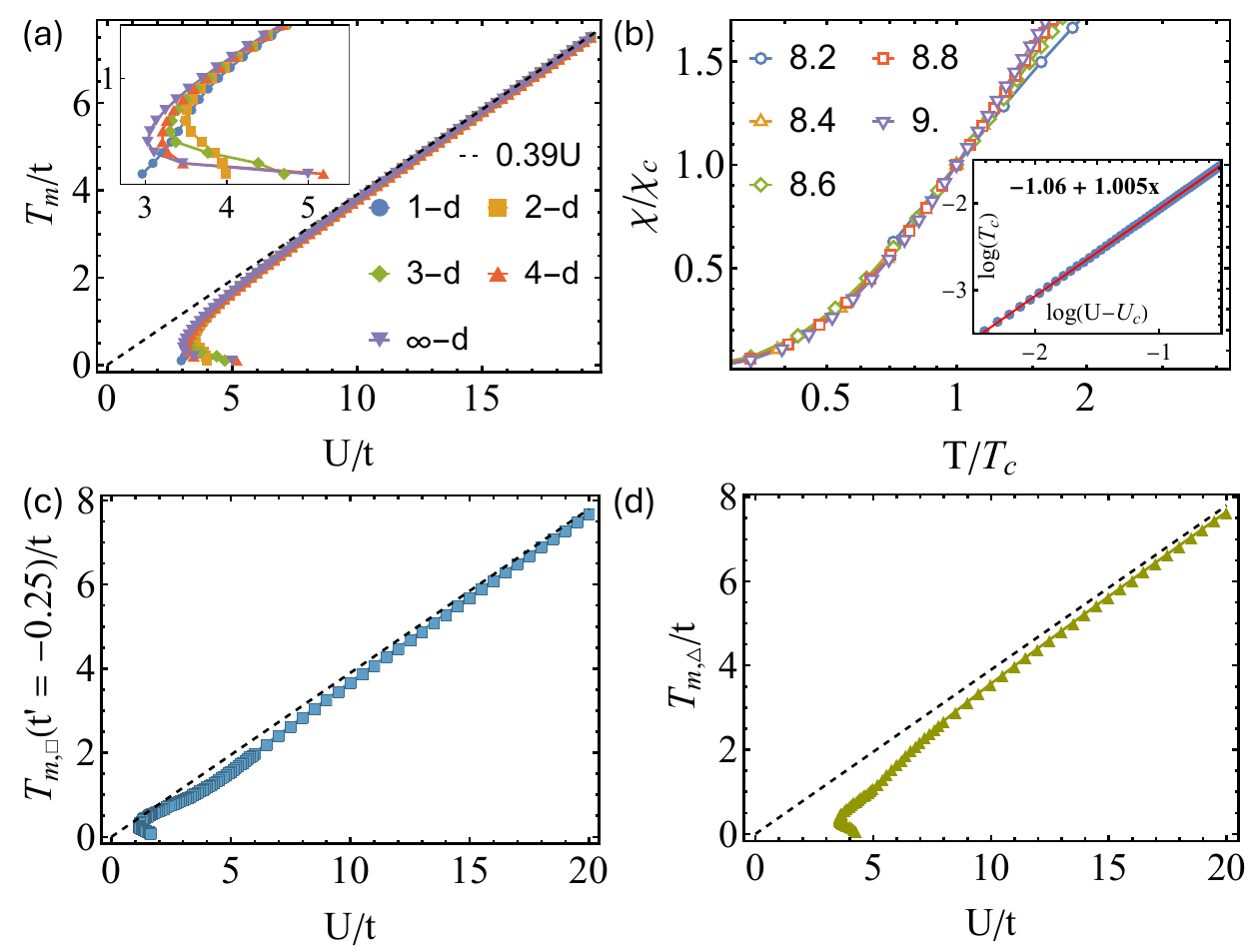}
    \caption{ (a) The widom lines: $T_m$, the temperature at which the compressibility $\chi$ has an extremum, as a function of $U$ for  band HK model for $d=1, 2, 3, 4$ and $d=\infty$ dimensions. 
    (b) Scaled compressibility $\chi/\chi_c$ versus scaled temperature $T/T_c$ for $U>U_c$ for band HK model on $2$D square lattice. 
    (c) The widom line for $4$-MMHK model on a 2-d square lattice with $t'/t=-0.25$. 
    (d) The widom line for $4$-MMHK model on a 2-d triangular lattice. 
     }
    \label{fig:widom}
\end{figure}

The metal-insulator transition is defined by the extrema of the compressibility as a function of temperature. Fig.~\ref{fig:widom}(a) plots $T_m$ as a function of $U$ for HK models in $d=1, 2, 3, 4$ and $d=\infty$ dimensions. These melting curves correspond to the Widom line\cite{stanley}.  Physically, the Widom line corresponds to a line emanating from the critical point that demarcates the maxima of thermophysical quantities on either side of the transition.  In all dimensions, the asymptotic ratio of $T_m/U\approx0.39$ is observed regardless of the underlying lattice structure, as predicted from the high-temperature analysis discussed above.  Since this ratio is determined entirely by the potential energy, the $\mathbb{Z}_2$ universality is abundantly manifest.

In dimensions greater than $1$, $T_m$ has multiple solutions around the Mott transition. Consider the cut in Fig.~\ref{fig:widom}(a) at $U$ slightly less than $U_c$. The first solution $T_m^{(1)}$ above 0 represents a local minimum: the system is metallic for $T<T_m^{(1)}$ but becomes an insulator for $T>T_m^{(1)}$. This anomalous temperature-driven Mott transition is also observed in DMFT\cite{dmft}.
This is sometimes recognized as the smoking gun of a first-order transition. However, there is no instability in the free energy. The extrema of the compressibility are third derivatives of the free energy. Thus, multiple solutions for $T_m$ could represent a possible third-order transition.  However, the turn-on of the compressibility above and below the transition confirms that it is indeed second order. Similar behavior of the Widom line is observed in MMHK, see Fig.~\ref{fig:widom}(c,d), but not in band insulators. The robust shape and universal high temperature ratio of $T_m/U\approx0.39$ of the Widom lines are signatures for the Mottness fixed point.

The compressibility data around the Mott transition collapse onto a universal scaling behavior for the metal-insulator transition. We find that the compressibility can be rescaled as illustrated in Fig.~\ref{fig:widom}(b). To rescale the low temperature behavior, we choose $T_c$ so $\left.\frac{\partial^2\chi}{\partial^2 T}\right|_{T=T_c}$ acquire its minimum value, which makes the low temperature inflection points overlap, and $\chi_c=\chi(T_c)$ is the corresponding compressibility. The scaling exponent $\alpha$ defined as $T_c\propto|U-U_c|^\alpha$ fits to the value $\alpha\approx1$. Similar scalings are also achieved for MMHK models with $U>U_c$.
\begin{figure}[t]
    \centering
    \includegraphics[width=\linewidth]{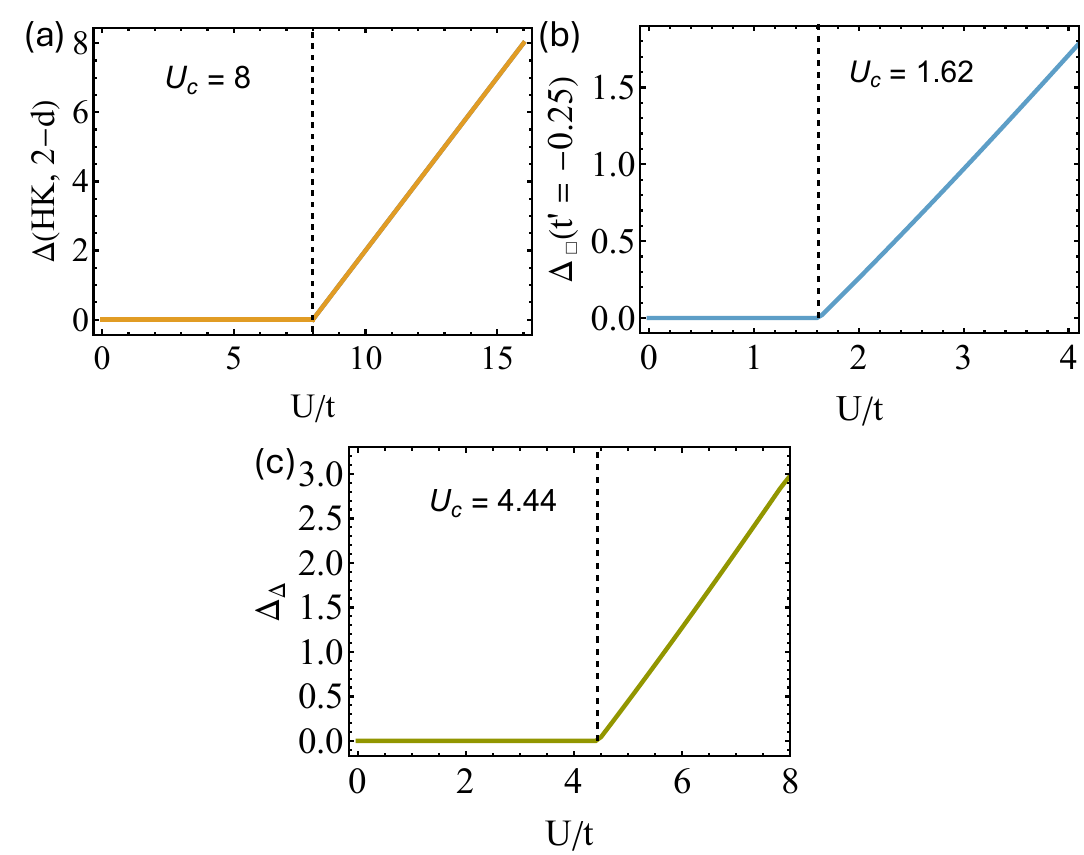}
    \caption{ The gap size $\Delta$ as a function of $U$ for (a) band HK model on $2$-d square lattice with $t'=0$, (b) $4$-MMHK model on square lattice with $t'/t=-0.25$, (c) $4$-MMHK model on triangular lattice. The Mott transition is marked by the vertical dashed line.}
    \label{fig:gap}
\end{figure}

It is the evolution of the insulating gap as a function of $U$ that dictates the critical scaling. As shown in Figs.~\ref{fig:gap} (a-c), the insulating gap increases linearly with $U-U_c$ for both the band HK and various MMHK models. In the insulating phase, the gap $\Delta$ serves as the energy scale governing low-energy and low-temperature properties. This is corroborated by the collapse of the compressibility curves when plotted as $\chi/\chi_c$ vs $T/T_c$, where $T_c\propto \Delta\propto(U-U_c)$. The resulting relation naturally gives the scaling exponent $\alpha=1$. This universal exponent holds for both band and MMHK models.

\begin{figure}[t]
    \centering
    \includegraphics[width=\linewidth]{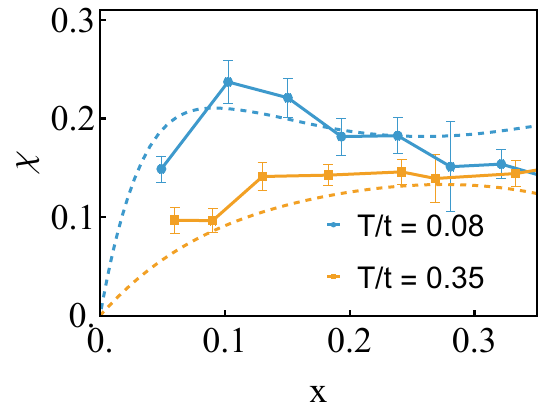}
    \caption{Dots: The quantum simulator compressibility as a function of doping level in $2$D square lattice Hubbard model with $U/t=7.00(4)$ (permission from \cite{coldatom}), connected by solid lines. Dashed lines: computed compressibility in the doped 4-MMHK model on $2$D square lattice at $U/t = 7$. The yellow lines are high-temperature $(T /t = 0.35)$ data, the blue lines are data at low temperature $(T /t = 0.08)$.}
    \label{fig:doped}
\end{figure}

The extension to finite doping reveals how the $\mathbb{Z}_2$ universality evolves away from half-filling, particularly in relation to the pseudogap phenomena\cite{zxarpes,pseudo1} which exhibits a clear departure from Fermi liquid theory. Recent progress in cold-atom quantum simulators on the Hubbard model\cite{coldatom} reveals a crossover between a normal metal and the pseudogapped metal defined by the  doping level at which the compressibility achieves a maximum at a fixed temperature. A quick check with the 4-MMHK compressibility as shown in Fig.~\ref{fig:doped} resembles some crucial features achieved in the quantum simulator (see Fig. 2 (b) of ref.\cite{coldatom}): 
\begin{enumerate}
    \item The high temperature $(T /t = 0.35)$ compressibility saturates to a steady value, which is also confirmed by DQMC\cite{coldatom}.
    \item The low temperature $(T /t = 0.12)$ compressibility strongly peaks at $x\approx0.1$.  Such low temperatures pose no impediment for MMHK method, in contrast to DQMC which faces the sign problem.  MMHK is based on an enumeration of the eigenstates and hence can achieve even $T=0$ without any sign problem.  The consistency of the location and the value of the compressibility peak between the quantum simulator and 4-MMHK result signifies the power of the $\mathbb{Z}_2$ universality.  On general grounds, a peak in $\chi$ as a function of doping arises because both $x=0$ and $x=1$ are insulating phases with vanishing compressibility.  Hence, there must be a peak at some doping level.  Since the peak in $\chi$ moves to lower doping as the temperature decreases, it remains an open question whether the peak tracks the pseudogap line\cite{pseudo1,pseudo2}.  This will be explored in a forthcoming publication. 
\end{enumerate}

\emph{Discussion}- This work advances a unified perspective on the Mott metal-insulator transition by identifying the compressibility $\chi$—rather than double occupancy—as a robust order parameter that exhibits singular behavior across the transition. Analysis of a family of exactly solvable models, from the local-in-momentum band HK model to its MMHK extensions, demonstrates that the transition is characterized by the breaking of an inherent $\mathbb{Z}_2$ symmetry. This symmetry, rooted in the structure of Fermi liquids, underlies the emergence of Hubbard bands and the opening of the Mott gap without reliance on a mean-field order parameter.

The results establish several universal features of the Mott insulator. First, the Widom line---defined by the locus of extrema in $\chi(T)$---exhibits asymptotic behavior $T_m \propto U$ deep in the insulating phase, with a universal prefactor of approximately $0.39$, independent of dimension, lattice structure, and the details of model. This reflects the dominance of interaction effects and confirms the $\mathbb{Z}_2$ universality class. Second, the insulating gap $\Delta$ scales linearly with $U - U_c$, leading to a critical scaling exponent $\alpha = 1$ for the characteristic temperature $T_c \propto |U - U_c|$. The collapse of $\chi/\chi_c$ versus $T/T_c$ further validates the universality of this scaling across all models considered.

These findings challenge the conventional view that double occupancy serves as the order parameter for the Mott transition. While double occupancy decreases continuously with $U$ in models with dynamical mixing, the compressibility clearly distinguishes metallic from insulating phases. This reconciliation helps resolve discrepancies between DMFT\cite{Rohringer2018,Park2008}, which predicts a first-order transition in infinite dimensions, and recent numerical studies of 2-d Hubbard models\cite{Leblanc2015,yanagi2014}, which show continuous behavior. 

Several promising directions emerge from this work. Besides the pseudogap phase, the strange-metal behavior at finite doping extends to extraordinarily high temperatures. Instead of pure charge correlation, the current correlation in MMHK requires advanced response techniques. Furthermore, connecting these results to experimental probes---such as scanning tunneling microscopy or optical conductivity---in materials such as the vanadates, organic salts, or cuprates, could provide direct tests of the Widom line and critical scaling reported here.

\textbf{Acknowledgements} 
We thank Lev Kendrick of Martin Greiner's group for extensive discussions at LT30 and for sending us his compressibility data in Fig. 4 from their cold-atom quantum simulator.
This work was supported by the Center for Quantum Sensing and Quantum Materials, a DOE Energy Frontier Research Center, grant DE-SC0021238 (P. M. and P. W. P.). JZ and PWP also acknowledges NSF DMR-2111379 for partial funding of the HK work which led to these results. P.M. was also supported by the Gordon and Betty Moore Foundation’s EPiQS Initiative through grant GBMF 8691.

\section*{Appendix}\label{appendix}
\subsection{The double occupancy as an order parameter?}
\begin{figure}[h]
    \centering
    \includegraphics[width=\linewidth]{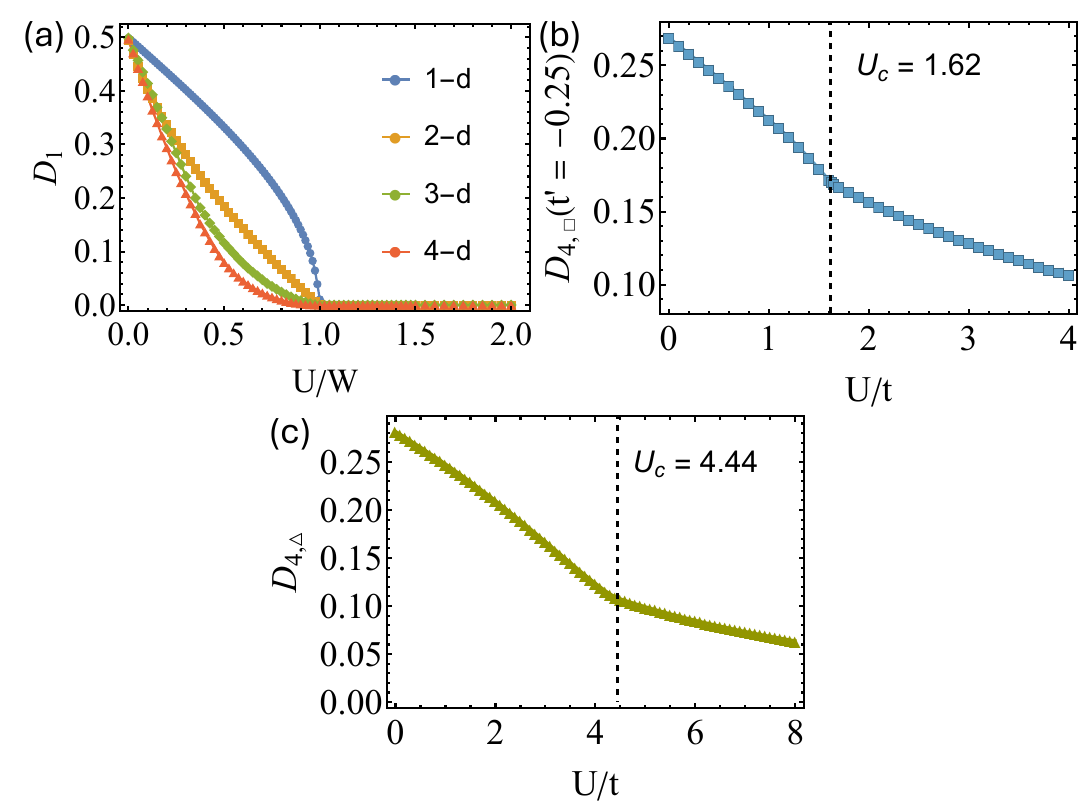}
    \caption{ (a) double occupancy $D$ as a function of $U$ for band HK model in $d=1$, $2$, $3$, $4$, and $\infty$. 
    (b) Double occupancy $D_4$ as a function of $U$ in $4$-MMHK model on 2-d square lattice with $t'/t=-0.25$.
    (c) Double occupancy $D_4$ as a function of $U$ in $4$-MMHK model on 2-d triangular lattice.}
    \label{fig:do_v_U}
\end{figure}

The double occupancy is proportional to the average interaction strength\cite{twisted}:
\begin{equation}
D_n=\sum_{\mathbf{k}\in \rm{rBZ}_n}\sum_\alpha \langle n_{\mathbf{k}\alpha\uparrow}n_{\mathbf{k}\alpha\downarrow}\rangle,
\end{equation} 
with explicit dependence on $n$.
A naive interpretation of Mott insulators is that the interaction eliminates double occupancy in the system, as depicted by the atomic limit Hubbard model or the band HK model, Fig. \ref{fig:do_v_U}. However, once the dynamics are restored in terms of momentum-scattering, the double occupancy no longer vanishes in the Mott insulating phases as shown in Fig.~\ref{fig:do_v_U}(b) and (c), but only in the $U\rightarrow\infty$ limit. Since there is no qualitative difference in double occupancy between the insulator and the metal once dynamics are included, it is not a valid indicator of the Mott transition.  We conclude then that the low-temperature compressibility instead of double occupancy should be viewed as the order parameter of the Mott transition.

\subsection{The compressibility in metalic states}
Metals are compressible as they are characterized by filling surfaces where the single-particle Green function supports a simple pole with a finite residue\cite{ppfixedp}.  As a calibration, we start with the non-interacting case. Setting $U=0$ in Eq. (\ref{eq:comp}) leads immediately to 
\begin{equation}
\begin{split}
    \chi(T)&=\sum_{k\in BZ}\frac{\beta}{1+\cosh(\beta\xi_k)}\\
    &=\int d\xi D(\xi)\frac{\beta}{1+\cosh(\beta\xi)},
\end{split}
\end{equation}
where we have changed the summation over the Brillouin zone into an integral over energy, and $D(\xi)$ is the density of states measured from the Fermi surface.  As the integrand dies off exponentially for states away from the Fermi surface, we Taylor expand the density of states $D(\xi+\mu)$ around the Fermi surface as 
$D(\xi)=D(0)+D'(0)\xi+\frac{1}{2}D''(0)\xi^2+O(\xi^3)$. The leading contribution yields
\begin{equation}
\chi(T)=2D(0)+\frac{2\pi^2}{3}D''(0)T^2+O((\beta\xi)^3),
\end{equation}
which demonstrates the result that the low-temperature compressibility is related to the density of states on the Fermi surface. 


To investigate the explicit dependence on $U$ in a Mott insulator, we consider the low-temperature behavior with $\beta U\gg1$. The compressibility
\begin{equation}
\begin{split}
    \chi(T)&=\sum_{k\in BZ}\frac{2\beta}{Z_k^2}\left(e^{-\beta\xi_k}+2e^{-\beta(2\xi_k+U)}+e^{-\beta(3\xi_k+U)}\right)\\
    &\approx\int d\xi D(\xi)\frac{2\beta e^{\beta\xi}}{\left(2+e^{\beta\xi}\right)^2}+O\left(e^{-\beta U}\right)\\
    &\quad+\int d\xi D(\xi+U)\frac{2\beta e^{-\beta(\xi+U)}}{\left(2+e^{-\beta(\xi+U)}\right)^2},
\end{split}
\end{equation}
has contributions from both the lower  ($\xi=0$) and  upper ($\xi+U=0$) filling surfaces. The term $e^{-\beta(2\xi_k+U)}/Z_k^2$ is suppressed by $\exp(-\beta U)$ compared to the other terms, thus marked as $O(e^{-\beta U})$. The total compressibility is therefore
\begin{equation}
    \begin{split}
    \chi(T)&=D(0)+D(U)+\log2\left(D'(0)-D'(U)\right)T\\
    &+\left(\frac{\pi^2}{3}+\log(2)^2\right)\left(D''(0)+D''(U)\right)T^2+O((\beta\xi)^3).\\
\end{split}
\end{equation}
Still, the low temperature compressibility is proportional to the total density of states at both filling surfaces.

\bibliography{mottbib}
\end{document}